\documentclass[fleqn,letters,usenatbib]{mnras}

\usepackage[T1]{fontenc}
\usepackage{ae,aecompl}

\usepackage{newtxtext,newtxmath}

\usepackage{graphicx}	
\usepackage{amsmath}	
\usepackage{amssymb}	

\title[Effective temperatures of metal-poor benchmarks]{Accurate effective temperatures of the metal-poor benchmark stars HD\,140283, HD\,122563 and HD\,103095 from CHARA interferometry}

\author[I. Karovicova et al.]{I. Karovicova,$^{1}$\thanks{E-mail: karovicova@uni-heidelberg.de (IK)}
           T. R. White,$^{2}$
           T. Nordlander,$^{3}$        
           K. Lind,$^{4,5}$
           L. Casagrande,$^{3}$ \newauthor
           M. J. Ireland,$^{3}$                  
           D. Huber,$^{6,7,8,2}$
           O. Creevey,$^{9}$
           D. Mourard,$^{9}$
           G. H. Schaefer,$^{10}$
           G. Gilmore,$^{11}$ \newauthor
           A. Chiavassa,$^{9}$  
           M. Wittkowski,$^{12}$ 
           P. Jofr\'{e},$^{13}$      
           U. Heiter,$^{5}$
           F. Th\'{e}venin,$^{9}$ 
           and M. Asplund$^{3}$
\\
$^{1}$ Zentrum f\"{u}r Astronomie der Universit\"{a}t Heidelberg, Landessternwarte,
   K\"{o}nigstuhl 12, 69117 Heidelberg \\
$^{2}$ Stellar Astrophysics Centre, Department of Physics and Astronomy, Aarhus University, Ny Munkegade 120, DK-8000 Aarhus C, Denmark\\
$^{3}$  Research School of Astronomy \& Astrophysics, Australian National University, Canberra, ACT 2611, Australia\\
$^{4}$ Max Planck Institute f\"ur Astronomy, K\"{o}nigstuhl, Heidelberg, 69117, Germany, \\
$^{5}$ Observational Astrophysics, Department of Physics and Astronomy, Uppsala University, Box 516, 75120 Uppsala, Sweden\\
$^{6}$ Institute for Astronomy, University of Hawai`i, 2680 Woodlawn Drive, Honolulu, HI 96822, USA \\
$^{7}$ Sydney Institute for Astronomy (SIfA), School of Physics, University of Sydney, NSW 2006, Australia \\
$^{8}$ SETI Institute, 189 Bernardo Avenue, Mountain View, CA 94043, USA \\
$^{9}$ Universit\'{e} de La C\^{o}te d'Azur, OCA, Laboratoire Lagrange CNRS, BP. 4229, 06304, Nice Cedex, France \\
$^{10}$ Department of Physics and Astronomy,
Georgia State University
P.O. Box 5060,
Atlanta, GA 30302-5060 \\
$^{11}$ Institute of Astronomy, University of Cambridge, Madingley Road, Cambridge, CB3 0HA, United Kingdom \\
$^{12}$ European Southern Observatory, Karl-Schwarzschild-Str. 2, 85748, Garching bei M\"unchen, Germany \\
$^{13}$ N\'ucleo de Astronom\'ia, Universidad Diego Portales, Av. Ej\'ercito 441, Santiago, Chile\\
}

\date{Accepted XXX. Received YYY; in original form ZZZ}

\pubyear{2015}

\begin{document}
\label{firstpage}
\pagerange{\pageref{firstpage}--\pageref{lastpage}}
\maketitle

\begin{abstract}
{Large stellar surveys of the Milky Way require validation with reference to a set of ``benchmark'' stars whose fundamental properties are well-determined. For metal-poor benchmark stars, disagreement 
between spectroscopic and interferometric effective temperatures has called the reliability of the temperature scale into question.
   We present new interferometric measurements of three metal-poor benchmark stars, HD\,140283, HD\,122563, and HD\,103095, from which we determine their effective temperatures.
  The angular sizes of all the stars were determined from observations with the PAVO beam combiner at visible wavelengths at the CHARA array, with additional observations of HD\,103095 made with the VEGA instrument, also at the CHARA array. Together with photometrically derived bolometric fluxes, the angular diameters give a direct measurement of the effective temperature. 
   For HD\,140283 we find $\theta_\mathrm{LD}$=0.324$\pm$0.005\,mas, $T_\mathrm{eff}$=5787$\pm$48\,K;
	for HD\,122563, $\theta_\mathrm{LD}$=0.926$\pm$0.011\,mas, $T_\mathrm{eff}$=4636$\pm$37\,K;
and for HD\,103095 $\theta_\mathrm{LD}$=0.595$\pm$0.007\,mas, $T_\mathrm{eff}$=5140$\pm$49\,K. Our temperatures for HD\,140283 and HD\,103095 are hotter than the previous interferometric measurements by 253\,K
and 322\,K, respectively.  
   We find good agreement between our temperatures and recent spectroscopic and photometric estimates.
   We conclude some previous interferometric measurements have been affected by systematic uncertainties larger than their quoted errors.
}
   
\end{abstract}




\section{Introduction}

Understanding the stellar populations of the Galaxy relies upon precise determination of fundamental stellar properties. 
One of the main challenges in the determination of global parameters of stars is an accurate estimate of the effective temperature, $T_\mathrm{eff}$.
Temperatures are commonly derived from spectroscopy or photometry, however, neither of these techniques provide a direct determination of this parameter.
The most direct and, in comparison to spectroscopy or photometry, nearly model-independent measurements of $T_\mathrm{eff}$ come from the combination of interferometric measurements of the angular diameter, $\theta$, with a determination of the bolometric flux, $F_\mathrm{bol}$, given by
\begin{equation}
     T_\mathrm{eff}  =   \left( \frac{4F_\mathrm{bol}}{\sigma\theta^2}\right)^{1/4},
\end{equation} 
where $\sigma$ is the Stefan-Boltzmann constant.


Current observational constraints limit precise interferometric measurements of the angular diameter $\theta$ to relatively bright ($V < 8$\,mag) stars with $\theta \gtrsim 0.3$\,mas. Consequently, measurements are limited to only a few metal-poor stars, typically close to these limits. In particular, angular diameters have been measured by \citet{Creevey12,Creevey15} for the brightest and most nearby metal-poor G-type subgiant HD 140283, K-type red giant branch star HD 122563, and K-type dwarf HD 103095. However, some differences have been found between inteferometric temperatures and those derived from spectroscopy and photometry that are difficult to reconcile, particularly for HD\,140283 and HD\,103095 \citep[see][and references therein for a detailed discussion]{Heiter15}. \citet{Heiter15} recommended that both these stars should not be used as temperature standards until these differences can be resolved.


As these stars are currently the only metal-poor benchmark stars \citep{Jofre14,Heiter15} selected for the large stellar survey conducted by the Gaia mission \citep{gaia16} as well as the supporting Gaia-ESO spectroscopic survey \citep{Gilmore12, Randich13}, and are used as standard stars in a large number of spectroscopic studies, it is of the utmost importance that these issues are resolved. In this study, we therefore present new interferometric observations at visible wavelengths of these three metal-poor stars.

\section{Observations}
\subsection{Interferometric observations and data reduction}

We observed the three stars using 
the PAVO beam combiner at the 
CHARA array at Mt. Wilson
Observatory, California \citep{tenBrummelaar05}.
PAVO is a pupil-plane beam combiner operating
between $\sim$\,600--900\,nm, with a spectral dispersion of each scan typically producing visibility measurements in 27 independent wavelength channels \citep{Ireland2008}.
The limiting magnitude of the PAVO instrument is $R$\,$\sim$\,7.5\,mag (8\,mag in ideal conditions). 
We observed the stars using baselines between 156.3\,m
and 313.6\,m. The stars were observed between 2014 Apr 8 and 2017 June 16. We also obtained VEGA \citep{vega} observations of HD103095 in March and May 2017. VEGA is a dispersed fringes beam combiner. 
For these observations, we defined a spectral band of 15nm or 20nm wide, around the central wavelength of the observations (either 700 or 720nm). Our observations are summarized in Table~\ref{table:2}. 

\begin{figure}
   \centering
   \includegraphics[width=\columnwidth]{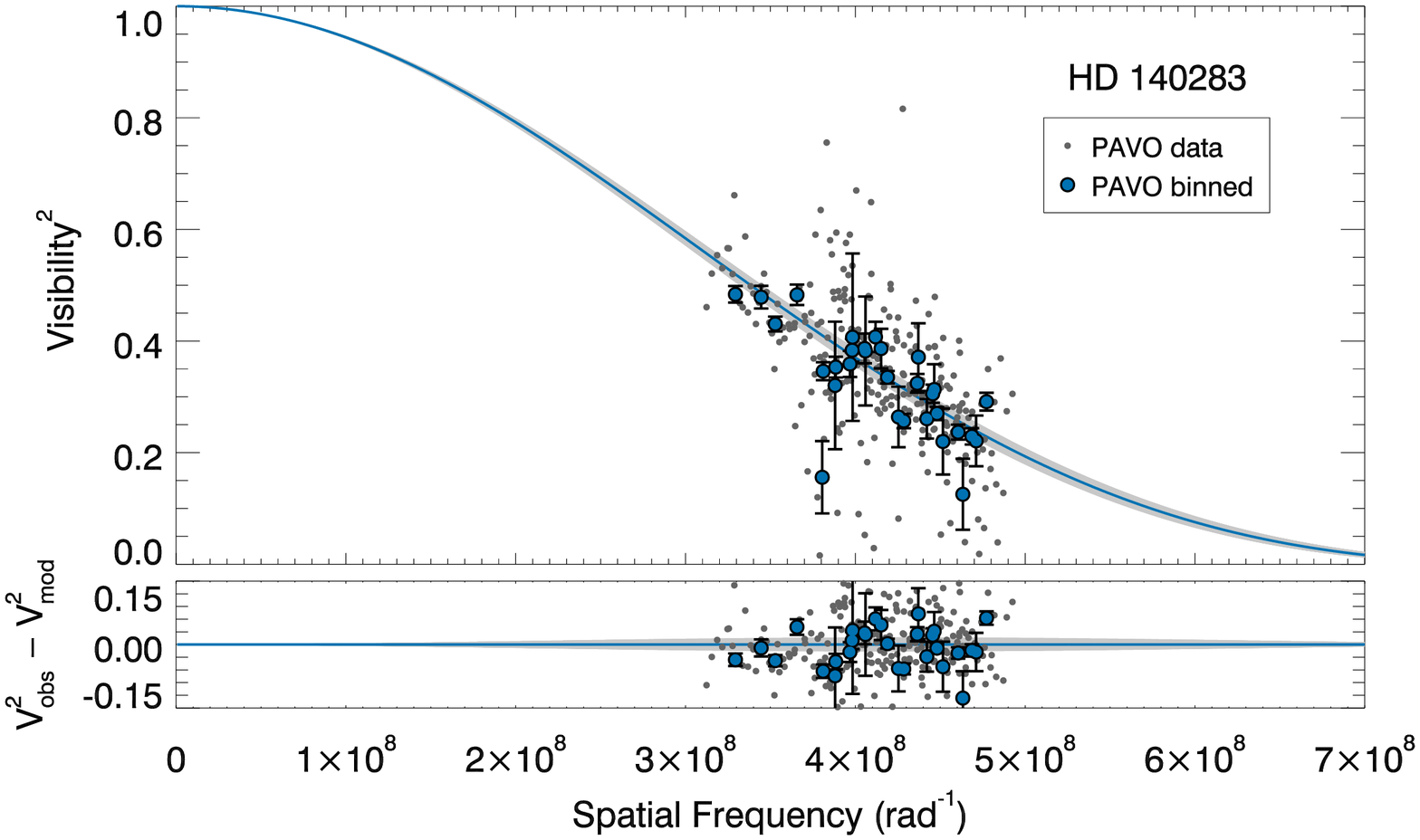}
   \includegraphics[width=\columnwidth]{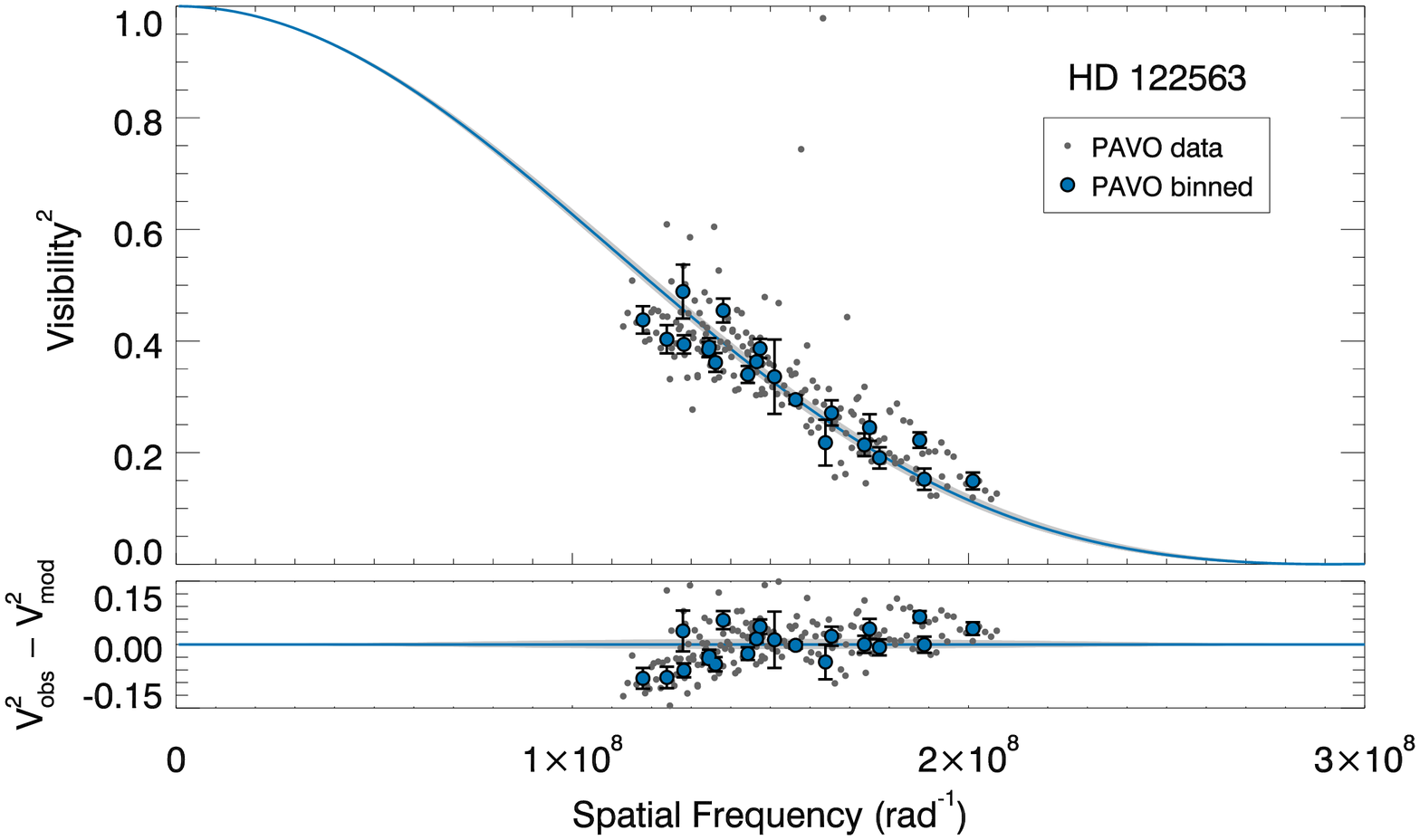}
   \includegraphics[width=\columnwidth]{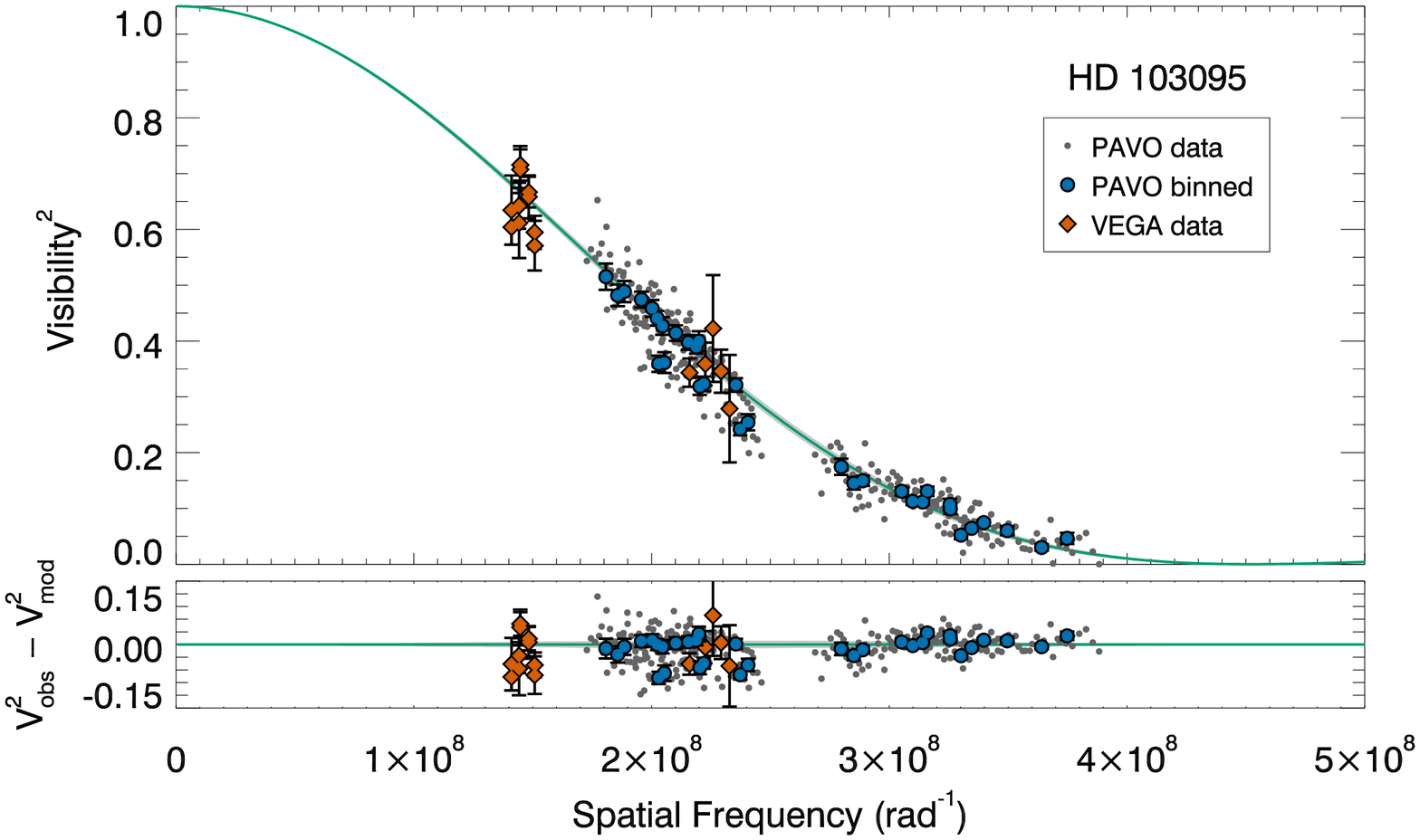}
      \caption{Squared visibility versus spatial frequency for HD\,140283, HD\,122563, and HD\,103095. The grey points are PAVO measurements. For clarity, we show weighted averages of the PAVO measurements over nine wavelength channels as blue circles. The blue line shows the fitted limb-darkened model to the PAVO data, with the light grey-shaded region indicating the 1-$\sigma$ uncertainties. For HD\,103095, measurements were also made with the VEGA beam combiner (orange diamonds), with the best-fitting model to the combined data indicated by the green line. The lower panels show the residuals from the fit.}
         \label{103095}
   \end{figure}

The raw data were reduced using the PAVO reduction software, which has been well-tested 
and used in multiple studies \citep{Bazot11,Derekas11,Huber12,Maestro13}.
To monitor the interferometric transfer function, we observed a set of calibration stars.
Calibrators, selected from the Hipparcos catalogue \citep{hipparcos}, were chosen to be likely unresolved sources and located close on the sky to the science target. Calibrators were
observed immediately before and after the science target.
The angular diameters of the calibrators were determined using the $V-K$ relation of \citet{boyajian14} and corrected for limb-darkening to determine the uniform disc diameter in $R$ band. We use $V$-band magnitudes from the Tycho-2 catalogue \citep{Hoeg2000} converted into the Johnson system using the calibration by \citet{bessell00}, and $K$-band magnitudes from the Two Micron All Sky Survey \citep[2MASS;][]{Skrutskie2006}. We use the dust map of \citet{green15} to estimate the reddening, and adopt the reddening law of \citet{odonnell94}. 
The diameter for the VEGA calibrator is obtained from the JMMC Stellar Diameters Catalog \citep[JSDC;][]{Chelli16}.
Details of the calibrators are summarized in Table~\ref{table:3}.
The calibrated squared visibility measurements of our three targets are shown in Fig.~\ref{103095} as a function of spatial frequency. 

HD\,140283 is a particularly difficult target to observe due to its small size, southern declination, and it being relatively faint. The observations made during 2017 were particularly challenging due to a temporarily reduced sensitivity of the CHARA Array while an adaptive optics system was installed and not yet operational. While the 2017 observations display an increased scatter in visibility, the measured diameter is consistent across all observing seasons. 

\begin{table}\footnotesize
\caption{Log of interferometric observations. }            
\label{table:2}      
\centering                          
\begin{tabular}{l l c c c c}      
\hline\hline            
HD & UT date & Combiner & Baseline$^a$ & Scans & Cal.$^b$ \\   
\hline
   140283 & 2014 Apr 8 & PAVO & E1W1 & 4 & jkl\\ 
             & 2015 Apr 4  & PAVO & S1W1 & 2 & jk\\
             & 2017 June 16 & PAVO & E1W1 & 4 & ik\\
   122563 & 2017 Mar 3 & PAVO  & E2W2 & 3 & ehi \\
             & 2017 June 9 & PAVO & E2W2 & 2 & gi \\
             & 2017 June 10 & PAVO & E2W2 & 2 & fi \\
   103095 & 2015 May 2 & PAVO  & E2W2 & 3 & ab \\
             & 2017 Mar 3  & PAVO & E2W2 & 3 & ad \\
             &             & PAVO & E2W1 & 2 & ad \\          
             & 2017 Mar 4  & PAVO & E1W2 & 3 & abd \\      
             & 2017 Mar 13 & VEGA & W1W2 & 3 & c \\
             & 2017 Mar 14 & VEGA & W1W2 & 3 & c \\  
             & 2017 May 5  & VEGA & S2W2 & 3 & c \\
\hline                                 
\end{tabular}
\flushleft $^{a}$ The baselines used have the following lengths: W1W2, 107.93\,m; E2W2, 156.27\,m; S2W2, 177.45\,m; E1W2, 221.85\,m; E2W1, 251.34\,m; S1W1, 278.50\,m; E1W1, 313.57\,m. 
\newline $^{b}$ Refer to Table~\ref{table:3} for details of the calibrators used.
\end{table}

   
\begin{table}\small
\caption{Calibrator stars used for interferometric observations.}            
\label{table:3}      
\centering                          
\begin{tabular}{l l c c c c c}      
\hline\hline            
HD & Sp. T. & $V$ & $K$ & $E(B-V)$ & $\theta_{\mathrm{UD},R}$ & ID\\
   &          & (mag) & (mag) & (mag) & (mas) & \\
\hline                      
 99002   & F0    & 6.93 & 6.28 & 0.008 & 0.201(10) & a \\  
 103288  & F0    & 7.00 & 6.22 & 0.006 & 0.211(11) & b \\
 103928	 & A9V   & 6.42 & 5.60 & 0.002 & 0.282(7)  & c \\
 107053  & A5V   & 6.68 & 6.02 & 0.004 & 0.226(11) & d \\   
 120448  & A0    & 6.78 & 6.52 & 0.017 & 0.169(8)  & e \\  
 120934  & A1V   & 6.10 & 5.96 & 0.007 & 0.216(11) & f \\  
 121996  & A0Vs  & 5.76 & 5.70 & 0.029 & 0.238(12) & g \\
 122365  & A2V   & 5.98 & 5.70 & 0.007 & 0.248(12) & h \\
 128481  & A0    & 6.98 & 6.79 & 0.007 & 0.149(7)  & i \\    
 139909  & B9.5V & 6.86 & 6.54 & 0.110 & 0.165(8)  & j \\    
 143259  & B9V   & 6.64 & 6.28 & 0.107 & 0.187(9)  & k \\  
 146214  & A1V   & 7.49 & 7.10 & 0.012 & 0.132(7)  & l \\
\hline                                 
\end{tabular}
\end{table}


\begin{table*}\small
\caption{Angular diameters and limb-darkening coefficients.}   \label{table:4}      
\centering                          
\begin{tabular}{lccccccccc}      
\hline\hline            
Star &  Combiner & $\theta_\mathrm{UD}$ (mas) & \multicolumn{2}{c}{Linear limb darkening$^a$} &  \multicolumn{5}{c}{4-term limb darkening$^a$} \\   
     &             &                 &    $u$    &    $\theta_\mathrm{LD}$ (mas)   & $a_1$ & $a_2$ & $a_3$ & $a_4$         & $\theta_\mathrm{LD}$ (mas) \\
\hline                      
   HD\,140283 & PAVO & 0.312 $\pm$ 0.006 & 0.550$\pm$0.009 & 0.327$\pm$0.005 & 1.62$\pm$0.11 & $-$2.11$\pm$0.30 & \enspace\,2.01$\pm$0.32 & $-$0.68$\pm$0.12 & 0.324 $\pm$ 0.005 \\    
   HD\,122563 & PAVO & 0.882 $\pm$ 0.010 & 0.632$\pm$0.006 & 0.937$\pm$0.011 & 1.26$\pm$0.09 & $-$1.20$\pm$0.13 & \enspace\,1.25$\pm$0.13 & $-$0.43$\pm$0.05 & 0.926 $\pm$ 0.011 \\
   HD\,103095 & PAVO & 0.565 $\pm$ 0.005 & 0.631$\pm$0.008 & 0.601$\pm$0.005 & 0.49$\pm$0.11 & \enspace\,0.51$\pm$0.14 & $-$0.30$\pm$0.12 & \enspace\,0.06$\pm$0.05 & 0.595 $\pm$ 0.007 \\
     & VEGA & 0.582 $\pm$ 0.008 &  & 0.617$\pm$0.009 &  &  &  &  & 0.611 $\pm$ 0.009 \\
     & PAVO+VEGA & 0.566 $\pm$ 0.005 &  & 0.602$\pm$0.005 &  &  &  &  & 0.596 $\pm$ 0.007 \\
\hline                                 
\end{tabular}
\flushleft $^{a}$ Limb-darkening coefficients derived from the grid of \citet{magic15}; see text for details.
\end{table*}

\begin{table}\small
\caption{Observed ($\theta_\mathrm{LD}$) and derived ($F_\mathrm{bol}$, $T_\mathrm{eff}$, $L$, $R$) stellar parameters. }.        
\label{table:5}      
\centering                          
\begin{tabular}{c c c c c}      
\hline\hline            
Parameters&HD\,140283 &   HD\,122563       & HD\,103095  \\   

\hline     
   m$_V$ (mag) & 7.21  & 6.19 & 6.45 \\
   
   m$_R$ (mag) & 6.63  & 5.37 & 5.80 \\

   $\pi$ (mas) & $17.15 \pm 0.14$$^a$   & $4.22 \pm 0.35$$^b$  & $109.99 \pm 0.41$$^b$ \\
   
   
   $F_\mathrm{bol}$ (erg s$^{-1}$ & 3.93 $\pm$ 0.05  & 13.20 $\pm$ 0.29 & 8.22 $\pm$ 0.25 \\ 
   
   cm$^{-2}$10$^{-8}$)&   &  &  \\
   
   $\theta_\mathrm{LD}$ (mas) & 0.324 $\pm$ 0.005  & 0.926 $\pm$ 0.011 & 0.595 $\pm$ 0.007 \\
   
   $T_\mathrm{eff}$ (K) & 5787 $\pm$ 48  & 4636 $\pm$ 37 & 5140 $\pm$ 49 \\    
   
   
   
   $L$ (L$_\odot$) & 4.82 $\pm$ 0.27  & 232 $\pm$ 37 & 0.21 $\pm$ 0.01 \\
   
   $R$ (R$_\odot$) & 2.04$\pm$0.04  & 23.7$\pm$2.0 & 0.586$\pm$0.007 \\
\hline                                 
\end{tabular}
\flushleft $^a$ \citet{Bond13},
 $^b$ \citet{vanLeeuwen07}
\end{table}   
      
\subsection{Modelling of limb-darkened angular diameters}
\label{models}
Resolving small-scale structure, such as limb darkening, requires measurements in the sidelobes of the visibility function. The low contrast of the interference fringes in the sidelobes and the need for the star to be over-resolved limits such measurements to bright stars with large angular sizes \citep[e.g.][]{kervella17}. Our measurements are therefore degenerate between a uniformly-illuminated disc and a limb-darkened disc. Input from stellar model atmospheres is required to determine an appropriate amount of limb-darkening to infer the true angular diameters of these stars. Limb-darkening (LD) laws, which parametrize the intensity variation with a small number of coefficients, are commonly employed, with grids of coefficients calculated for various model atmospheres, limb-darkening laws and photometric filters \citep[e.g.][]{claret11}.

The use of the linear limb-darkening law is ubiquitous in interferometric studies due to its simple form and the ease with which different studies can be compared. However, it has long been known that the linear law does not fully describe the intensity distribution across the stellar disc in either observations or models \citep[e.g.][]{klinglesmith70}, and higher-order limb-darkening laws have been developed. To achieve a more faithful representation of the model atmosphere we used a four-term non-linear limb-darkening law \citep{claret00}. For ease of comparison to other interferometric studies, we have also fitted each star with a uniform disc model and a linearly limb-darkened disc model.

For a generalized polynomial limb-darkening law, 
\begin{equation}
\frac{I\left(\mu\right)}{I\left(1\right)} = \sum_{k} c_k \, \mu^k, \label{eqn:polyld}
\end{equation}
with $\mu = \cos\gamma$, and $\gamma$ is the angle between the line of sight and the emergent intensity,
the visibility is given by \citep{quirrenbach96}
\begin{equation}
V\left(x\right) = \left[\sum_{k}\frac{c_k}{k+2}\right]^{-1} \sum_{k} c_k\, 2^{k/2}\, \Gamma\left(\frac{k}{2}+1\right)\frac{J_{k/2+1}\left(x\right)}{x^{\left(k/2+1\right)}}, 
\end{equation}
where $x=\pi B \theta \lambda^{-1}$, with $B$ the projected baseline, $\theta$ the angular diameter, and $\lambda$ the wavelength of observation, $\Gamma(z)$ is the gamma function, and $J_n(x)$ is the $n$th-order Bessel function of the first kind. The quantity $B\lambda^{-1}$ is the spatial frequency.

Limb-darkening coefficients in the $R$-band were determined from limb-darkening tables computed by \citet{magic15} from the \textsc{Stagger}-grid, a set of state-of-the art 3D radiation-hydrodynamical model atmospheres that are computed from first principles. We generated initial values of the limb-darkening coefficients by interpolating the grid of coefficients to the spectroscopic atmospheric parameters adopted from the literature \citep{Bensby14, Bergemann12, Ramirez13}. For HD\,140283 the initial values were $T_\mathrm{eff}$ = 5658\,K, $\log(g)$ = 3.6 and [Fe/H] = $-$2.62\,dex; for HD\,122563 they were  $T_\mathrm{eff}$ = 4665$\pm$80\,K, $\log(g)$ = 1.64$\pm$0.16 and [Fe/H] = $-$2.51\,dex; for HD\,103095 they were $T_\mathrm{eff}$ = 5149$\pm$70\,K, $\log(g)$ = 4.71$\pm$0.3 and [Fe/H] = $-$1.27$\pm$0.2\,dex. We estimate uncertainties $T_\mathrm{eff, \sigma}$ = 90\,K, $\log(g)_\sigma$ = 0.3 and [Fe/H]$_\sigma$ = 0.2. The influence on LD coefficients from deviations from the initial spectroscopic values are negligible. 


Uncertainties in the spectroscopic parameters were used to generate 10\,000 realizations of the limb-darkening coefficients to estimate their uncertainties, and, in the case of the four-term law, their correlations. Additionally, we determined 1D limb darkening coefficients based on \citet{claret11} 
using the same spectroscopic values as for the 3D simulations (Table ~\ref{table:4}). The results 
based on 1D and 3D models are consistent within the uncertainties and the uncertainties are in both cases similar. We used both 1D and 3D modelling for comparison, however, the 3D simulations are important for a complete characterization of the stellar surface properties. The 3D-based limb-darkening is better supported by the observed solar centre-to-limb variation \citep{pereira13} and exoplanet transit light curves \citep{hayek12}. Further details will be given in a following paper (Karovicova et al. in prep).

The fit to the interferometric measurements was performed following the procedure described by \citet{Derekas11}, estimating angular diameter uncertainties by performing Monte Carlo simulations that take into account the uncertainties in the visibility measurements, wavelength calibration (0.5 per cent), calibrator sizes and the limb-darkening coefficients. For HD\,103095 we performed fits to the PAVO and VEGA data, both separately and combined.

The new interferometric $T_\mathrm{eff}$ was calculated using the derived angular diameter and bolometric flux (see Sect 2.3). The limb-darkening coefficients were then redetermined using the new $T_\mathrm{eff}$, with the process iterated until the results converged. The final angular diameter measurements are listed in Table~\ref{table:4}, and the fits of the 3D limb-darkened diameters are shown in Figure~\ref{103095}.


\subsection{Bolometric flux}

\begin{figure}
   \centering
   \includegraphics[width=\columnwidth]{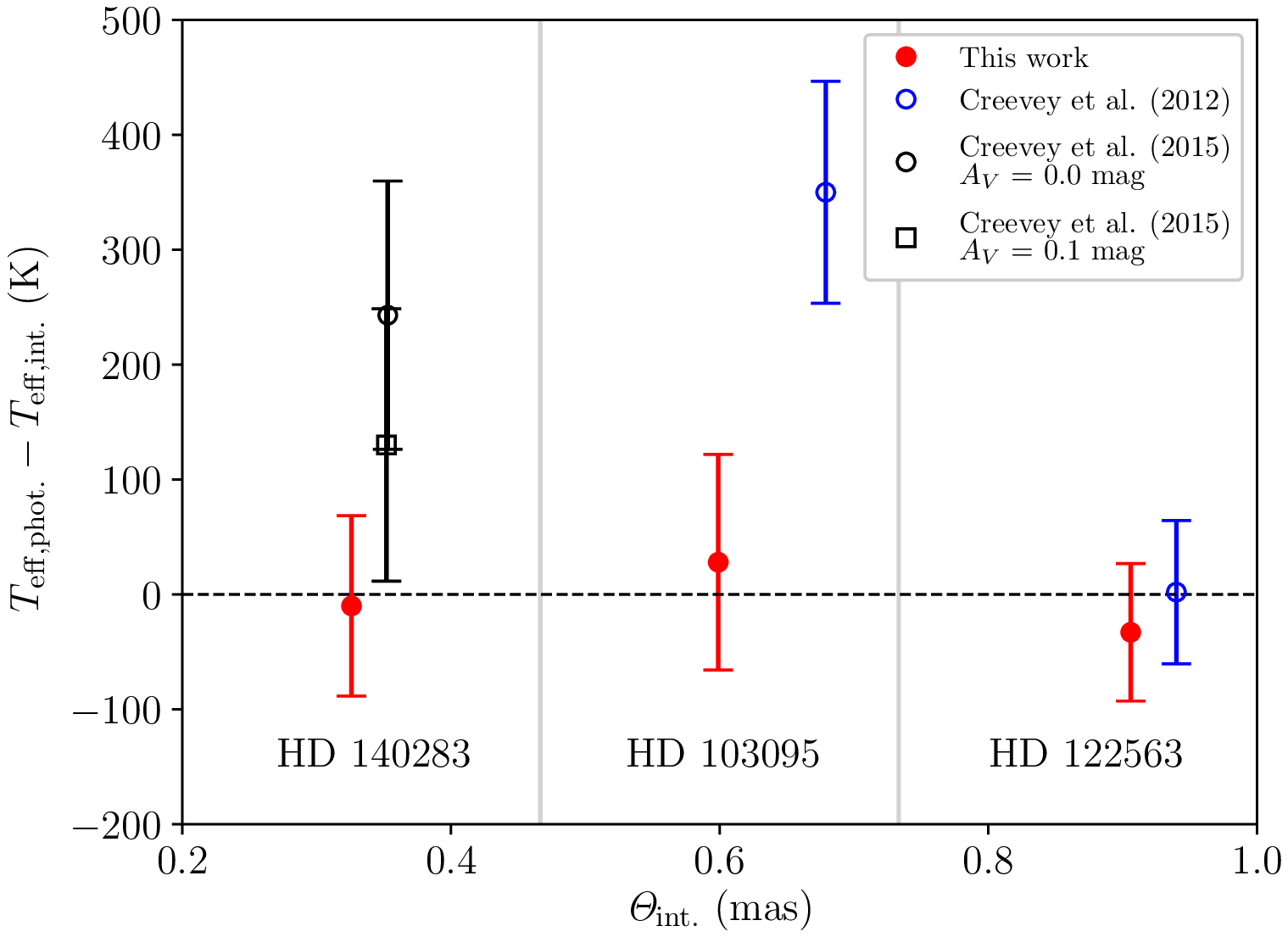}
      \caption{Filled red circles are effective temperatures $T_\mathrm{eff,phot}$ - $T_\mathrm{eff,int.}$ versus
interferometric measurements.  $T_\mathrm{eff,phot}$ \citet{Casagrande10} for HD 140283,
\citet{Casagrande14} for HD\,122563 and \citet{Casagrande11} for HD\,103095. Blue
open circles represent measurements from \citet{Creevey12} for HD 103095 and HD
122563, and black open symbols from \citet{Creevey15} a circle for HD 140283 and an open square for A$_V$ = 0.1 mag.
      }
         \label{Teff}
   \end{figure}

Bolometric fluxes for HD\,140283 and HD\,122563 were derived using the InfraRed
Flux Method (IRFM) as described in \citet{Casagrande10, Casagrande14}. Only
HD\,140283 has 2MASS photometry of sufficiently high quality for this implementation of the IRFM,
although for HD\,122563 
%
we circumvented the problem by converting Johnson $JHK$ photometry into the 2MASS 
system\footnote{http://www.astro.caltech.edu/$\sim$jmc/2mass/v3/transformations
where Johnson has been converted using the 2MASS-Bessell \& Brett.} The IRFM depends very mildly on the adopted $\log(g)$ and
[Fe/H] of the stars. We iterated the IRFM with spectroscopic parameters until
reaching convergence in bolometric fluxes. For HD\,103095, we used bolometric
corrections in the Tycho2 $B_T V_T$ and Hipparcos $H_p$ system from \citet{Casagrande_VandenBerg14}. For this star, bolometric corrections were computed at the spectroscopic $T_\mathrm{eff}, \log(g)$ and [Fe/H]. We remark that the effective temperatures we derive from the IRFM, as well as the adopted spectroscopic $T_\mathrm{eff}$ for HD\,103095 are all in agreement with interferometric ones to within a few tens of degrees at most (Fig.~\ref{Teff}).
We adopted a reddening of zero for both HD\,103095 and HD\,140283 because of their vicinity ($\sim$10 and 60\,pc respectively, and hence within the local bubble \citep[e.g.][]{Leroy93,Lallement03}. Interstellar Na\,I\,D lines confirm the absence of reddening for HD\,140283 \citep{Melendez10}. HD\,122563 is the only star showing non-zero reddening, and we adopt $E(B-V)=0.003$\,mag, following \citet{Creevey12} and consistent with our preliminary analysis of interstellar Na\,I\,D lines (Karovicova et al. in prep.). A considerable higher value such as
$E(B-V)=0.01$\,mag for this star would increase the bolometric flux by 1\%, which is well within our adopted uncertainties.
%

\section{Results and discussion}

Our updated fundamental stellar parameters of the three metal-poor benchmark stars are given in Table~\ref{table:5}. Both the uncertainty in the angular diameter and the bolometric flux contribute significantly to the final uncertainty in the effective temperature. For HD\,140283 the angular diameter contributes $\sim$75\,per cent, for HD\,122563 they both contribute $\sim$50\,per cent, and for HD\,103095 the bolometric flux contributes $\sim$70\,per cent.

Our angular diameters for HD\,140283 and HD\,103095 are smaller than those determined in previous interferometric studies of these stars \citep{Creevey12,Creevey15}. For HD\,103095, our measurements with VEGA imply a slightly larger diameter (0.611$\pm$0.009\,mas) than is obtained from our PAVO measurements (0.595$\pm$0.007\,mas), however both are substantially smaller than previously obtained with the FLUOR and Classic beam combiners at the CHARA Array (0.679$\pm$0.015\,mas).

As a consequence of the smaller diameters, the inferred effective temperatures of these stars are higher. 
For HD\,140283 it is 253K (140K assuming non-negligible reddening A$_V$=0.1\,mag) higher, 
and for HD\,103095, $T_\mathrm{eff}$ is 322K higher, bringing the interferometric values into better agreement with spectroscopy and photometry. HD\,122563 is in a good agreement. In Fig.~\ref{Teff} we show the difference between recent photometric temperatures determined by \citet{Casagrande10,Casagrande11,Casagrande14} and interferometric measurements presented here and by \citet{Creevey12,Creevey15}.

The differences between our angular diameters and those determined by \citet{Creevey12,Creevey15} may be the result of systematic errors arising from the notoriously difficult calibration of interferometric observations. The system response of the interferometer is determined by measuring calibrator stars. However, errors in the predicted size of calibrators and atmospheric variability can lead to poor estimates of the system visibility. In an effort to minimize this, we observed multiple calibrators that were as small and as close as possible to the target over several observations on different nights. 

Comparisons between photometric and interferometric temperatures have noted that diameters measured in the $K'$ band with the Classic beam combiner at the CHARA Array with sizes $\lesssim$\,1\,mas appear to be systematically larger than expected \citep{Casagrande14}. This disagreement tends to increase with decreasing angular size, suggesting that systematic errors arise for near infrared observations when the stars are under-resolved. The differences we see with the previous measurements of these metal-poor benchmark stars is in the same direction and of similar magnitude to those noted by \citet{Casagrande14}, suggesting we are seeing the same systematic effect. By comparison, our temperatures are consistent with those determined photometrically, as shown in Fig.~\ref{Teff}. Our observations at visible wavelengths allow us a greater resolution to avoid these systematic errors. Some of the observations of HD\,103095 and HD\,122563 analysed by \citet{Creevey12} were CHARA Classic $K'$ band measurements, which may have contributed to the differences with our results.

The results presented here, in resolving the differences between the interferometric, spectroscopic and photometric temperatures of these important stars, now allow for their reliable use as benchmark stars for calibrating large stellar surveys.

\section*{acknowledgements}
      IK acknowledges support by the
      Deut\-sche For\-schungs\-ge\-mein\-schaft, DFG\/ project
      number KA4055 and by the European Science Foundation - GREAT Gaia Research for European Astronomy Training. This work was supported by Sonderforschungsbereich SFB 881
   ``The Milky Way System'' (subproject A3) of the DFG.
      This work is based upon observations obtained with the Georgia State University Center for High Angular Resolution Astronomy Array at Mount Wilson Observatory. The CHARA Array is supported by the National Science Foundation under Grants No. AST-1211929 and AST-1411654. Institutional support has been provided from the GSU College of Arts and Sciences and the GSU Office of the Vice President for Research and Economic Development.
      Funding for the Stellar Astrophysics Centre is provided by The Danish National Research Foundation. TRW acknowledges the support of the Villum Foundation (research grant 10118).
      MA, TN, LC, and DH gratefully acknowledge support from the Australian Research Council (grants DP150100250, FT160100402, and DE140101364, respectfully).
 DH also acknowledges support from the National Aeronautics and Space Administration under Grant NNX14AB92G issued through the Kepler Participating Scientist Program. GG acknowledges ERC grant 320360. KL acknowledges funds from the Alexander von Humboldt Foundation in the framework of the Sofja Kovalevskaja Award endowed by the Federal Ministry of Education and Research as well as funds from the Swedish Research Council (Grant $nr. 2015-00415_3)$ and Marie Sklodowska Curie Actions (Cofund Project INCA 600398). UH acknowledges support from the Swedish National Space Board (SNSB/Rymdstyrelsen).




\bibliographystyle{mnras}


\bsp	
\label{lastpage}
\end{document}